\documentclass[print]{ieeecolor}
\usepackage{generic}
\usepackage{cite}
\usepackage{amsmath,amssymb,amsfonts}
\usepackage{algorithmic}
\usepackage{graphicx}
\usepackage{algorithm,algorithmic}
\usepackage{hyperref}
\hypersetup{hidelinks=true}
\usepackage{textcomp}
\usepackage{array}
\def\BibTeX{{\rm B\kern-.05em{\sc i\kern-.025em b}\kern-.08em
T\kern-.1667em\lower.7ex\hbox{E}\kern-.125emX}}
\begin{document}
\noindent\fbox{%
    \parbox{\textwidth}{%
        \textbf{Notice:} This document is a preprint and has been accepted for publication in the \textit{IEEE Journal of Biomedical and Health Informatics}. The final, published version can be accessed using the following DOI: \href{http://dx.doi.org/10.1109/JBHI.2024.3385504}{10.1109/JBHI.2024.3385504}. Copyright for this article has been transferred to IEEE.
    }%
}
\title{Conditional Diffusion Models for \\ Semantic 3D Brain MRI Synthesis}
\author{Zolnamar Dorjsembe, Hsing-Kuo Pao, Sodtavilan Odonchimed, Furen Xiao
\thanks{
"This work was supported by the National Science and Technology Council, Taiwan [Grant No. 111-2221-E-002-049-MY3, 112-2221-E-011-111, NSTC 112-2634-F-011-002-MBK] and National Taiwan University Hospital [Grant No. 110-EDN03]."   (Corresponding author: Furen Xiao.)}
\thanks{Z. Dorjsembe and H. Pao are with the Department of Computer Science and Information Engineering, National Taiwan University of Science and Technology, Taipei 106, Taiwan (e-mail: d11115806@mail.ntust.edu.tw, pao@mail.ntust.edu.tw). }
\thanks{S. Odonchimed is with the Faculty of Engineering, The University of Tokyo, Tokyo 113-8654, Japan (e-mail: 1184142454@g.ecc.u-tokyo.ac.jp).}
\thanks{F. Xiao is with Institute of Medical Device and Imaging, National Taiwan University, College of Medicine, Taipei 100, Taiwan (e-mail: fxiao@ntu.edu.tw).}}

\maketitle

\begin{abstract}
Artificial intelligence (AI) in healthcare, especially in medical imaging, faces challenges due to data scarcity and privacy concerns. Addressing these, we introduce Med-DDPM, a diffusion model designed for 3D semantic brain MRI synthesis. This model effectively tackles data scarcity and privacy issues by integrating semantic conditioning. This involves the channel-wise concatenation of a conditioning image to the model input, enabling control in image generation. Med-DDPM demonstrates superior stability and performance compared to existing 3D brain imaging synthesis methods. It generates diverse, anatomically coherent images with high visual fidelity. In terms of dice score in the tumor segmentation task, Med-DDPM achieves 0.6207, close to the 0.6531 dice score of real images, and outperforms baseline models. Combined with real images, it further increases segmentation accuracy to 0.6675, showing the potential of our proposed method for data augmentation. This model represents the first use of a diffusion model in 3D semantic brain MRI synthesis, producing high-quality images. Its semantic conditioning feature also shows potential for image anonymization in biomedical imaging, addressing data and privacy issues. We provide the code and model weights for Med-DDPM on our GitHub repository (https://github.com/mobaidoctor/med-ddpm/) to support reproducibility.
\end{abstract}

\section{Introduction}
\label{sec:introduction}
\IEEEPARstart{D}{eep} learning has achieved remarkable progress in the medical field \cite{b02, b01, b03}. However, it faces challenges due to the scarcity and heterogeneity of medical data, the costs of annotation, and privacy concerns \cite{b2-1, b1-1, b1-2, b2-2}. To overcome these obstacles, generative models have emerged as a promising solution \cite{healthcare}, contributing to data augmentation \cite{image-augmentation}, image reconstruction \cite{image-reconstruction}, and privacy-preserving data anonymization \cite{privacy-preserving, anonymization}.

In the task of medical image synthesis, Generative Adversarial Networks (GANs) \cite{GAN} are a widely utilized method \cite{GANs_medical_image1, GANs_medical_image2}, but they face issues such as unstable training, mode collapse, and diminished gradients \cite{ganproblems}. Moreover, many GAN techniques primarily focus on two-dimensional (2D) images, which falls short of meeting the three-dimensional (3D) data requirements in medical imaging \cite{ganreview1, ganreview2}. A common approach generates 2D slices and stacks them to create 3D images, potentially causing spatial inconsistencies and neglecting 3D contextual information \cite{drawback2d_1, drawback2d_2}. Synthesizing meaningful, high-resolution 3D synthetic medical images, especially for complex organs like the brain, remains challenging.

Several studies have explored the synthesis of 3D medical images using GANs, employing latent vectors as inputs to generate images. G. Kwon et al. \cite{kwon2019} initially proposed the generation of 3D brain MRIs using an auto-encoding GAN. Building upon this, L. Sun et al. \cite{hagan} employed a hierarchical amortized GAN for high-resolution 3D medical image generation, with a focus on 3D thorax CT and brain MRI datasets. Extending this domain further, S. Hong et al. \cite{3dstylegan} proposed a method to adapt the StyleGAN2 model for 3D image synthesis in medical applications, specifically examining brain MR T1 images. These studies, however, concentrated primarily on unconditional image synthesis.

In contrast, semantic image synthesis offers a more controlled and customizable approach than unconditional generation, proving particularly beneficial in medical imaging for the precise synthesis of pathological images, such as accurately placing abnormal areas. Despite its potential, research in this domain remains limited, especially in the context of brain imaging. Noteworthy studies include those by A.B. Qasim et al. \cite{redgan}, who have employed the SPADE conditional generative network \cite{spade} for synthesizing new images from existing masks, focusing on 2D slice-wise brain image synthesis to preserve semantic information for segmentation tasks. Additionally, a significant contribution by H.-C. Shin. \cite{nvidia}, utilizing the original pix2pix conditional GAN model \cite{pix2pix} for label-to-MRI translation and MRI-to-label segmentation, stands out as a pioneering effort in 3D semantic image synthesis, being the only study to comprehensively address both 2D and 3D semantic image synthesis in the literature. However, training GAN models with the limited datasets, common in the medical domain, often leads to mode collapse, where the model generates similar data points, posing a significant challenge.

Diffusion models have recently emerged as a leading approach in generative modeling, achieving state-of-the-art results in generating high-quality, realistic images\cite{diffusion1, diffusion2}. This has led to a growing interest in exploring their potential applications in the field of medical imaging \cite{diffusioninmedicalimages}. 

Several studies have investigated the use of diffusion models for 3D medical image synthesis. Our previous work 3D-DDPM \cite{previouswork} was the first in the literature to apply diffusion models to 3D brain MRI unconditional synthesis, showing their superiority over GAN models. Following this, W. H. L. Pinaya et al. \cite{latent} advanced the field by employing Latent Diffusion Models for high-resolution 3D brain MRI synthesis, outperforming existing GAN models. However, despite these advancements, a significant research gap remains in semantic 3D medical image synthesis, with the majority of studies focusing on 2D images \cite{syndiff, diffusioninmedicalimages}.

Building on our previous work and aiming to fill the current research gap, this study focuses on improving Denoising Diffusion Probabilistic Models (DDPMs)\cite{ddpm} to address the challenges of limited annotated datasets and privacy concerns in the medical imaging domain. We propose a novel method, Med-DDPM, which incorporates segmentation masks into the diffusion process for pixel-level controllable 3D brain MRI synthesis. Our approach enables the generation of high-resolution, semantically guided 3D brain images and holds potential for extension to diverse image-to-image translation tasks within the medical domain.

To validate our approach, we conducted experiments using raw clinical brain MRI data without skull stripping. Specifically, we examined the impact of the synthesized images on the performance of the tumor segmentation task using a 3D U-Net model \cite{unet}. The results illustrate the superiority of our approach compared to GAN-based methods, showcasing a wide diversity of generated images and achieving results that closely align with real images in the segmentation task. Our proposed Med-DDPM demonstrates its remarkable effectiveness even with a small number of training images.

Furthermore, we conducted an additional experiment to validate the effectiveness of Med-DDPM, utilizing the brain-extracted MRI dataset from the BraTS2021 challenge\footnote{http://braintumorsegmentation.org/}. This experiment serves to showcase the remarkable capability of our proposed method in simultaneously generating all four modalities of MRI (T1, T1CE, T2, and Flair) from a segmentation mask.

Our contributions include: (1) Introducing Med-DDPM, a conditional diffusion model that utilizes pixel-level mask images for high-resolution 3D brain MRI synthesis. (2) Demonstrating empirical evidence, Med-DDPM significantly enhances segmentation model performance, bringing them closer to the accuracy achievable with real images. (3) Offering mask conditioning synthesis, enabling the generation of both normal and pathological whole-head MRIs with any size of abnormal areas, based on the given masks. Experimental results showcase the generation of diverse and high-quality images, suggesting the potential for Med-DDPM to serve as an advanced data augmentation and anonymization tool with further refinements. (4) Providing a publicly available synthetic dataset comprising brain pathological MR images with corresponding segmentation masks (
doi: 10.21227/3ej9-e459), alongside accessible code and model weights on our GitHub repository at https://github.com/mobaidoctor/med-ddpm/.

In summary, this research introduces a novel approach to semantic 3D brain MRI synthesis, emphasizing the potential of diffusion models in addressing challenges related to data scarcity and privacy preservation in the field of medical imaging.
\begin{figure*}[htbp]
\centering
\includegraphics[width=1.0\textwidth]{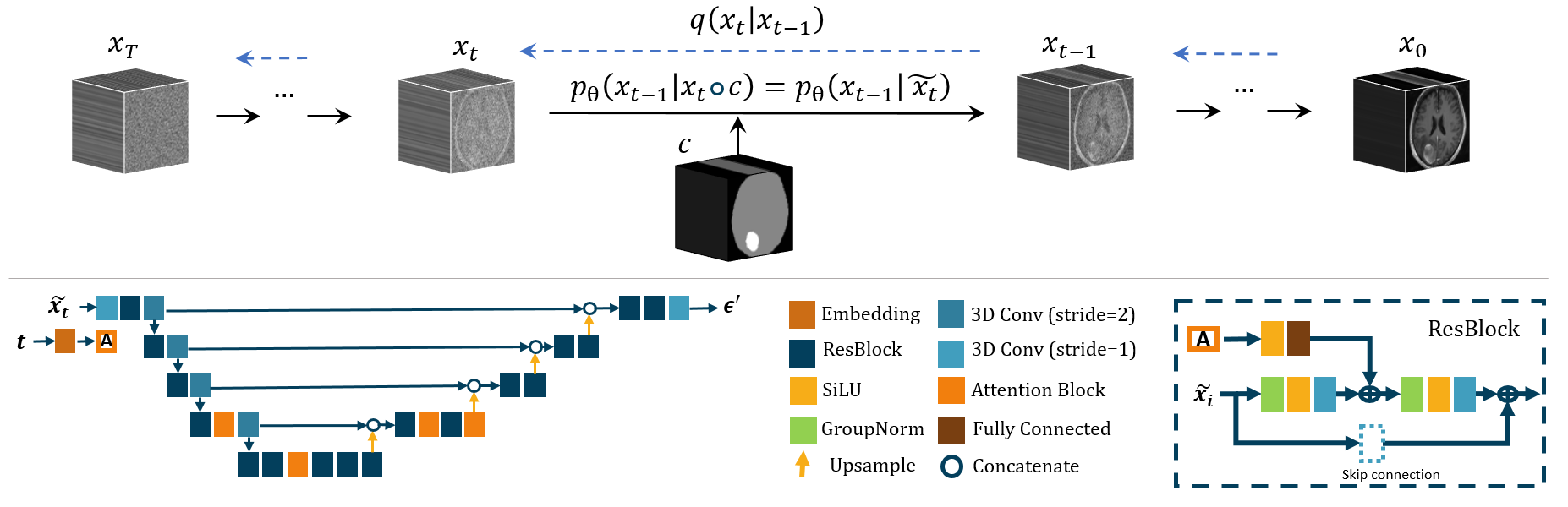}\vspace*{-2mm}
\caption{Architecture of the proposed method: The top row of the diagram demonstrates the conditioning mechanism of our approach, featuring the forward diffusion process $q(x_t | x_{t-1})$, and the denoising process $p_\theta (x_{t-1} | \tilde{x}_t)$. This process involves concatenating the conditioning mask $c$ with the input image $x_t$, resulting in the concatenated image $\tilde{x}_t$ utilized in the denoising process $p_\theta$. The bottom row presents an enhanced model architecture, adapted from \cite{previouswork}, providing a detailed view of the noise predictor U-Net model $\epsilon_\theta$ . This model predicts the noise $\epsilon'$, a critical component for the denoising process, as detailed in \eqref{eq:3}.}
\label{fig1}
\end{figure*}

\section{Method}
In this study, we extend our previous work, 3D-DDPM \cite{previouswork}, which adapted the vanilla DDPM model for generating 3D volumetric images. We further enhance this architecture to facilitate the synthesis of conditional 3D medical images, conditioned on segmentation masks. Initially, our approach was grounded on the original DDPM, as proposed in \cite{ddpm}.
The forward diffusion process $q$ adds small quantities of Gaussian noise ${\epsilon \sim \mathcal{N}(0,\mathbf{I})}$, defined by the variance schedule $\bar{\alpha_t}$, to an image sample $x_0$ from the training dataset at each timestep $t$ within a given number of timesteps $T$. The noisy sample $x_t$ is defined as follows for $1<t<=T$:
\begin{equation} x_t = \sqrt{\bar{\alpha_t}}x_0 + \sqrt{1-\bar{\alpha_t}}\epsilon.\label{eq:1}\end{equation}
To prevent sudden noise level fluctuations, we adopt a cosine noise schedule, as defined in \cite{improveddiffuion}:
\begin{equation} \bar{\alpha_t} = \frac{f(t)}{f(0)}, \ \
f(t)=\cos{(\frac{t/T+s}{1+s}\cdot \frac{\pi}{2})^2}, \label{eq:2} \end{equation}
Here, parameter $s$ is a small offset value that prevents the schedule from becoming exceedingly small as the timestep nears zero. In the reverse diffusion process $p_\theta$, we employed a modified 3D U-Net architecture, as described in our previous work \cite{previouswork}, to serve as our denoiser model. This modification involved replacing 2D components with their 3D equivalents to accommodate volumetric medical images and integrating embedding layers, Residual Blocks (ResBlocks), Sigmoid Linear Unit (SiLU) activation functions, group normalization, attention mechanisms, and fully connected layers to enhance model performance. 

Inspired by a super-resolution technique where a lower-resolution image is upsampled and concatenated to the generated image at each iteration, as demonstrated by J. Ho et al. \cite{inspiration}, we propose a straightforward, effective method that adapts this approach to modify the input image $x_t$ by channel-wise concatenating the segmentation mask. Contrary to the original DDPM, where $x_t$ is the only input, our model incorporates $\tilde{x_t}$, which includes an additional channel for the segmentation mask. This mask guides the generation process, enabling the synthesis of meaningful images, such as a pathological MRI of the brain with a tumor precisely positioned. The conditioning process is illustrated in Fig.~\ref{fig1}, where the segmentation mask $c$ is concatenated with the noisy image $x_t$ at each timestep $t$. Consequently, we also modify the number of input channels of denoiser model to match the number of segmentation mask labels. Fig.~\ref{fig1} illustrates the overall structure of our proposed method, including the model architecture adapted from \cite{previouswork}.

The training images and segmentation masks used in this study are single-channel volumetric images with three dimensions: width ($w$), height ($h$), and depth ($d$). The segmentation mask in the dataset consists of three class labels: 0 represents the background, 1 corresponds to the head area, and 2 indicates the tumor area. To prevent ordinal bias in model training due to the numeric class labels in the segmentation mask \cite{book}, we used one-hot encoding on the mask image. This process ignored the irrelevant background class label 0, creating a two-channel mask. In this mask, channel 0 represents the head area, and channel 1 indicates the tumor area. The channel-wise concatenation was then applied to combine the image and the mask, resulting in a concatenated image with three channels denoted as $\tilde{x_t}^{(3, w, d, h)} := x_t^{(1,w,d,h)} \oplus c^{2, w, d, h}.$ Throughout this paper, we refer to the concatenated image as $\tilde{x_t}$.
    
The denoising process, also known as the generative sampling process, $x_{t-1}\sim p_\theta (x_{t-1}\vert x_t)$, is formulated as follows:
\vspace{-2mm}
\begin{equation}
x_{t-1} = \frac{1}{\sqrt{\alpha_t}}(x_t - \frac{1-\alpha_t}{\sqrt{1-\bar{\alpha_t}}}\epsilon_\theta(\tilde{x_t}, t)) + \sigma_tz, \label{eq:3}
\end{equation}
where $z\sim \mathcal{N}(0,\mathbf{I}),\ \sigma_t=\sqrt{\beta_t}, \ \beta_t \in (0,1), \ \epsilon_\theta$ represents the trained noise predictor U-Net model. We present the complete training and sampling procedures in Algorithms~\ref{alg:training} and~\ref{alg:sample} respectively.
\vspace{-2mm}
\begin{algorithm}
\caption{Training}
\label{alg:training}
\begin{algorithmic}
\renewcommand{\algorithmicrequire}{\textbf{Input:}}
\REQUIRE $c$
\REPEAT
\STATE $x_0 \sim q(x_0)$
\STATE $t \sim \mathrm{Uniform}({\{1,..., T\}})$
\STATE $\epsilon \sim \mathcal{N} (0, \boldsymbol{I})$
\STATE $x_t = \sqrt{\bar{\alpha}_t}x_0+\sqrt{1-\bar{\alpha}}_t\epsilon$
\STATE $\tilde{x_t} = x_t \oplus c$
\STATE Take gradient descent step on \\
\hskip1.5em $\nabla_\theta L_{MAE}( \epsilon, \epsilon_\theta(\tilde{x_t}, t))$
\UNTIL{converged}
\end{algorithmic} 
\end{algorithm}
\vspace{-8mm}
\begin{algorithm}[H]
\caption{Sampling}
\label{alg:sample}
\begin{algorithmic}
\renewcommand{\algorithmicrequire}{\textbf{Input:}}
\REQUIRE $c$
\STATE sample $x_T \sim \mathcal{N} (0, \boldsymbol{I}) $,  $c$
\FOR{$t = T,...,1$}{ 
\STATE $\textbf{if} \  t > 1,\ \textbf{then} \ z \sim \mathcal{N} (0, I), \ \textbf{else}  \ z = 0 $
\STATE  $\tilde{x_t} = x_t \oplus c$\
\STATE   $\epsilon' = \epsilon_\theta(\tilde{x_t}, t)$ \
\STATE   $x_{t-1} = \frac{1}{\sqrt{\alpha_t}}\left( x_t - \frac{1-\alpha_t}{\sqrt{1-\bar{\alpha}_t}} \epsilon' \right) + \sigma_t z$\\
}
\ENDFOR
\STATE\algorithmicreturn{ $x_0$}
\end{algorithmic} 
\end{algorithm}

\subsection{Loss Function}
Pixel-wise losses such as $L1$ and $L2$ are commonly used in the literature of DDPM papers \cite{diffusion2}. In our study, we observed that the $L2$ loss (i.e., $L2=\mathbb{E}_{t,x_0,\epsilon}
[|\epsilon-\epsilon_\theta(\tilde{x_t},t)|^2])$ resulted in noisier images compared to the $L1$ loss (i.e., $L1=\mathbb{E}_{t,x_0,\epsilon} [|\epsilon-\epsilon_\theta(\tilde{x_t},t)|]).$ The $L2$ loss function, due to its computation of the squared difference between the estimated value and the target value, is sensitive to outliers. On the other hand, the $L1$ loss calculates the absolute differences between the estimated value and the target value, making it relatively less sensitive to outliers. Hence, in our main experiments, we utilize the $L1$ loss:
\begin{equation}
L_{MAE} = \frac{1}{n}\sum_{i=1}^n |\epsilon_i - \epsilon'_i|, \ \label{eq:4}
\end{equation}
where $\epsilon_i$ and $\epsilon'_i$
represent the pixels of the original noise added to the input and the predicted noise from the model (i.e., $\epsilon'=\epsilon_\theta (\tilde{x_t},t)$) respectively, $n$ is the total number of pixels ($n=w \cdot h \cdot d$).

\section{Experiments and Results}
\subsection{Datasets and Image Preprocessing} We used unnormalized clinical brain Magnetic Resonance (MR) images without skull stripping. Our evaluation was performed on the clinical stereotactic radiosurgery dataset \cite{b22}, which included 1688 contrast-enhanced T1-weighted (T1c) whole-head MR images and corresponding segmentation masks for various brain lesions. The dataset was obtained from patients undergoing Cyberknife radiosurgery at the National Taiwan University Hospital (NTUH).
For image preprocessing, we employed MRIPreprocessor\footnote{https://github.com/ReubenDo/MRIPreprocessor} for image registration and ensured consistent image dimensions by applying cropping and padding. The resulting dimensions were 192x192x192, which were then resized to 128x128x128 with a slice thickness of 1.5x1.5x1.5mm. To enhance the quality of our training data, we removed 188 outlier images that were highly distorted and exhibited strong artifacts. We then utilized only 1,500 images for our experiments. Following this, we performed intensity rescaling and normalized the image intensities to a range of [-1, 1]. The segmentation mask annotations included three classes: class 0 for the background, class 1 for the head, and class 2 for the tumor area.
To further evaluate our method, we conducted an additional experiment on multi-modality 3D brain MRI synthesis using the BraTS2021 challenge dataset\footnote{http://braintumorsegmentation.org/}. Details can be found in the "F. 3D Multimodal MRI Synthesis Experiment" subsection.

\subsection{Experiment Details} In this study, we primarily compared our work with the only existing study on 3D mask-to-image synthesis by H.-C. Shin et al. \cite{nvidia}. Additionally, we examined another study on semantic 2D brain MRI synthesis \cite{redgan} and a recent diffusion-based 2D image-to-image translation method \cite{syndiff}. However, their results were not satisfactory, and the quality further deteriorated when stacking them into 3D images, leading us to exclude them from the main comparison. We also utilized our previously developed 3D DiscoGAN architecture, which was adapted and modified from the method proposed by T. Kim et al. \cite{discogan}, serving as an additional baseline for comparison in semantic image synthesis. Furthermore, we evaluated the quality of our synthetic images by comparing them with the latest 3D brain MRI synthesis techniques \cite{hagan, 3dstylegan, kwon2019, latent}. Despite these methods being designed for unconditional synthesis, our goal was to benchmark the quality of our images against these recent advancements.

We trained our proposed model and the baseline GAN models using 1,292 images for 100,000 iterations with a batch size of 1. The evaluation of these models was performed on 208 testing images.

Our Med-DDPM model was trained using the L1 loss, the cosine noise schedule for 250 steps, a learning rate of $10^{-5}$ for the first 50,000 iterations, and $10^{-6}$ for the remaining 50,000 iterations. We utilized the Adam optimizer and refined the model parameters with an Exponential Moving Average (EMA) strategy, employing a decay factor of 0.995 to ensure stable and efficient training. The first layer of Med-DDPM consisted of 64 channels, and we incorporated an attention head at a resolution of 16. For the baseline conditional GAN models, 3D Pix2Pix and 3D DiscoGAN, we trained them using a combined loss of Mean Squared Error and L1 Losses, which yielded better results. These models were optimized with the Adam optimizer, with a learning rate set to $2 \times 10^{-4}$ and momentum decays of 0.5 and 0.999. For other unconditional 3D brain MRI synthesis models, we utilized their GitHub repositories and adhered to the specified original hyperparameters.

Additionally, to assess the effectiveness of synthetic images as a viable alternative to data augmentation, we conducted a comprehensive empirical study. Specifically, we utilized a simple 3D U-Net model \cite{unet} for the tumor segmentation task. The models were trained on synthetic data and compared against the performance of the 3D U-Net model trained on real data. The segmentation models underwent training for 100 epochs using Binary Cross-Entropy (BCE) loss with a batch size of 4. All models were trained on a Tesla V100–SXM2 32 GB GPU card.

\subsection{Evaluation Metric} We employed quantitative and qualitative measures to evaluate the synthetic images.

\textbf{Quantitative Evaluation:} The Fréchet Inception Distance (FID)\cite{b23} is a widely used evaluation metric in GAN synthesis, yet its application in assessing 3D medical images is limited due to its dependency on the Inception-v3 network \cite{b24} trained on a 2D image dataset. Following the implementation of the 3D-FID metric as described by \cite{hagan}, we adopted the same approach, utilizing a pre-trained 3D ResNet model\cite{b28} for feature extraction. Besides the 3D-FID metric, we utilized Mean Squared Error (MSE), Maximum Mean Discrepancy (MMD) score \cite{b29}, and Multi-Scale Structural Similarity Index Measure (MS-SSIM)\cite{b30} for quantitative evaluation.

\textbf{Qualitative Evaluation:} Two approaches were employed for the qualitative evaluation.

First, a visual evaluation was conducted with two neurosurgeons. The experts assessed a set of 40 high-resolution images, comprising an equal mix of 20 real and 20 synthetic images, with the objective of distinguishing them as real or fake. The images evaluated by the experts included samples from both a real image dataset and synthetic images generated by our proposed model. The experts made their classifications based solely on visual assessment. It is important to note that due to the easily identifiable non-realistic features of synthetic images generated by baseline GAN models, we did not proceed with further visual assessment of those GAN synthetic images.

Secondly, the tumor segmentation performance of a 3D U-Net model was evaluated to determine the effectiveness of synthetic images as a data augmentation technique. Various training scenarios were explored using synthetic dataset and real images.  We evaluated model performance using the Dice, Intersection over Union (IoU), Recall, and Precision scores. 

\begin{figure*}[htbp]
\vspace*{-5mm}
\centering
\includegraphics[width=1.0\textwidth]{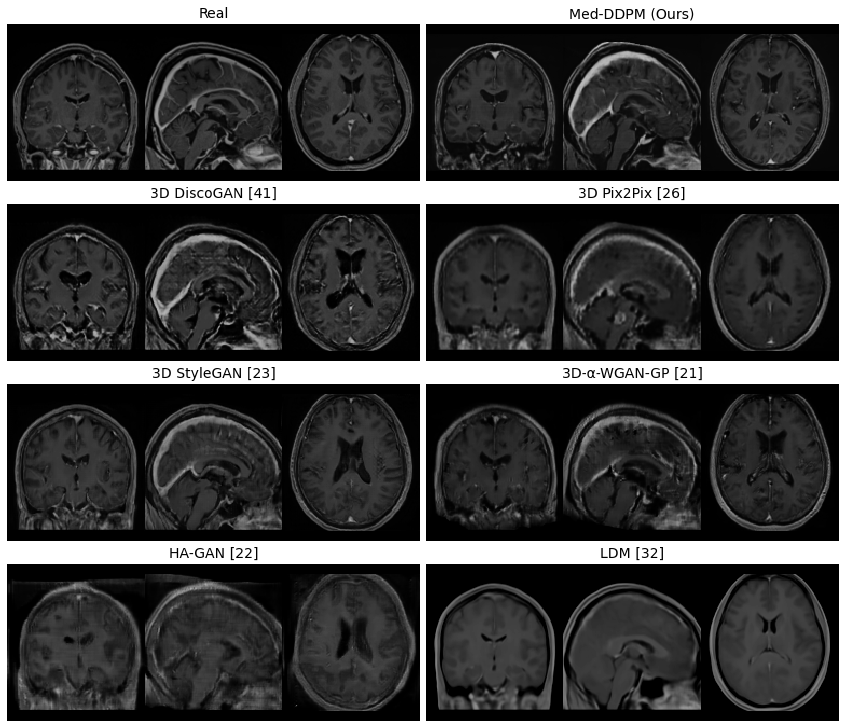}
\vspace{-6mm}
\caption{Comparison of overall quality in 3D brain MRI synthesis. This figure presents the quality comparison between real and synthetic 3D brain MRIs across coronal, sagittal, and axial slices. The first row displays a random real MRI sample alongside a synthetic sample from our proposed method. The second row presents random samples from baseline conditional synthesis methods. The final 2 rows showcase random samples from the latest unconditional synthesis methods, specifically designed for 3D brain MRI synthesis.}
\label{fig2}
\end{figure*}
\begin{figure*}[htbp]
\vspace{-4mm}
\centering
\includegraphics[width=0.95\textwidth]{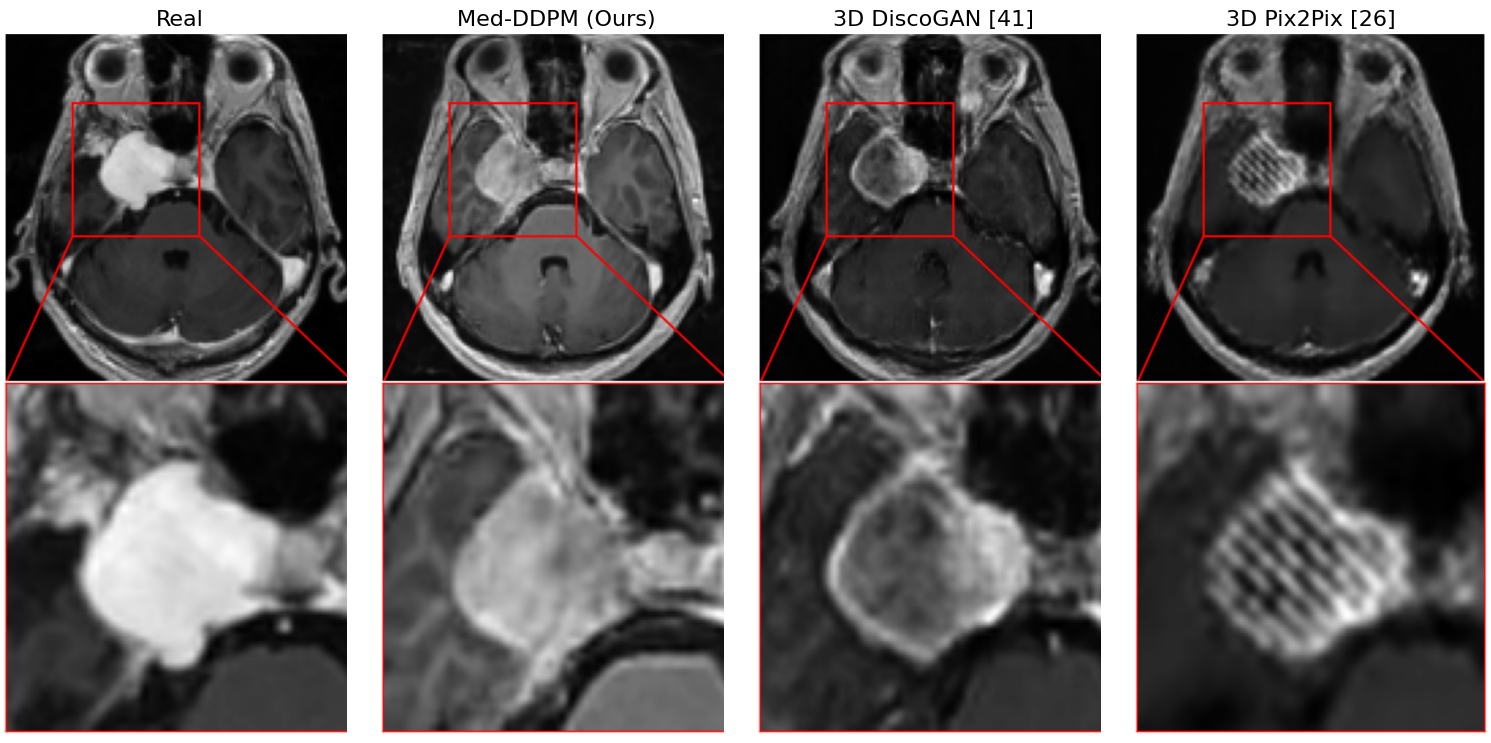}
\vspace{-2mm}
\caption{Zoomed visual comparison of tumor areas in real and generated samples (axial view slices). Med-DDPM and 3D DiscoGAN generate more realistic tumor parts with smoother edges and less artifacts. 3D Pix2Pix, on the other hand, has poor tumor synthesis, with strong artifacts that look unrealistic.}
\label{fig3}
\end{figure*}

\begin{figure*}[htbp]
\vspace{-2.5mm}
\centering
\includegraphics[width=0.95\textwidth]
{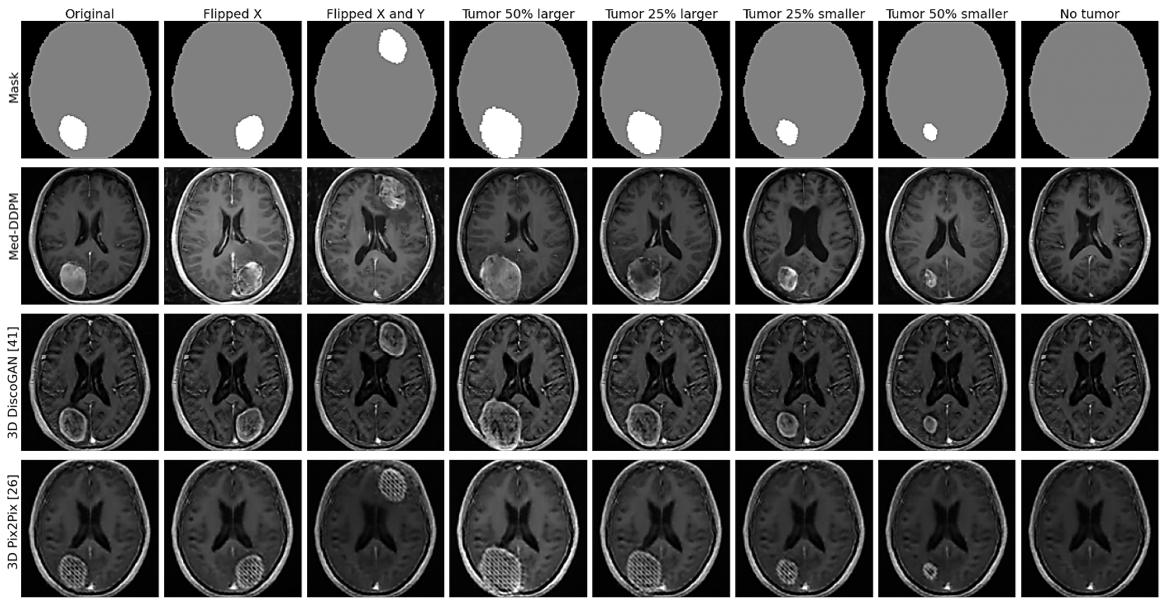}
\vspace{-2mm}
\caption{Comparison of synthetic images generated with manipulated masks (axial view slices). 3D DiscoGAN primarily captures the same brain features, with only slight variations in pixel intensities for tumor parts. 3D Pix2Pix also exhibits a similar limitation, highlighting the issue of mode collapse and the lack of diverse image generation in GAN models. In contrast, the proposed method, Med-DDPM, excels in synthesizing diverse images with strong variations.}
\label{fig4}
\end{figure*}
\begin{figure*}[htbp]
\vspace{-4mm}
\centering
\includegraphics[width=1.0\textwidth]{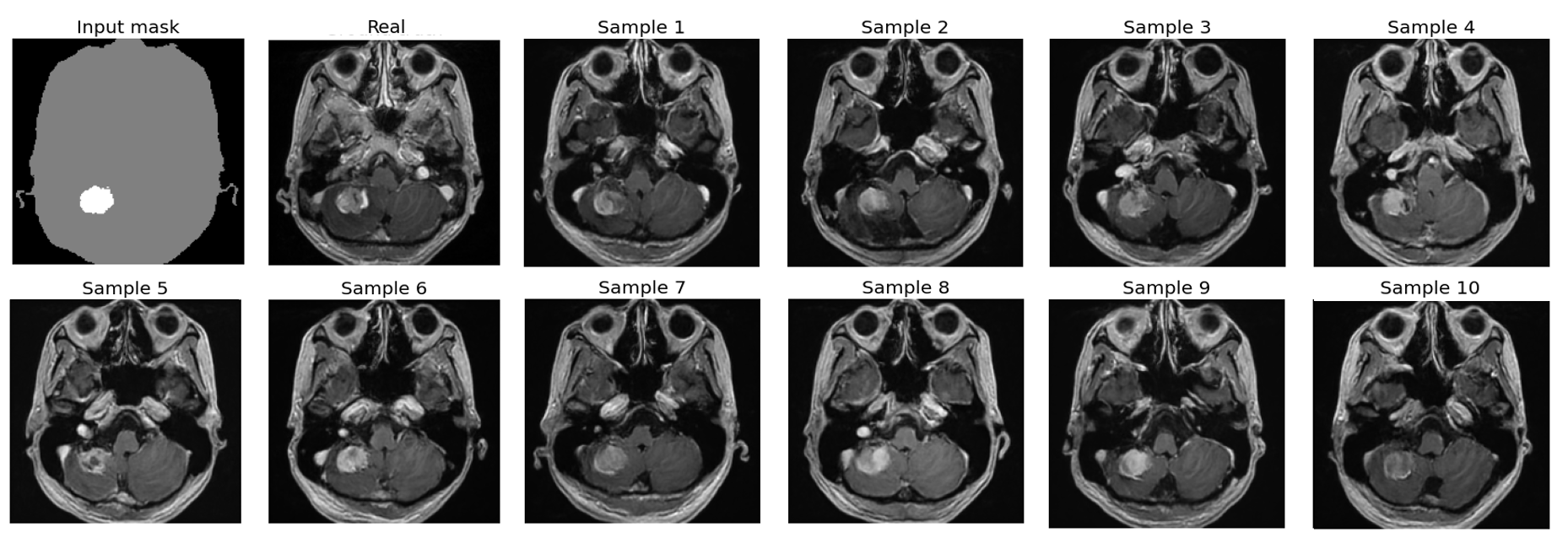}
\vspace{-6mm}
\caption{Center-cut axial slices of generated samples, showcasing the output diversity of Med-DDPM for a single input mask.}
\label{fig5}
\end{figure*}
\begin{figure*}[htbp]
\centering
\includegraphics[width=1.0\textwidth]{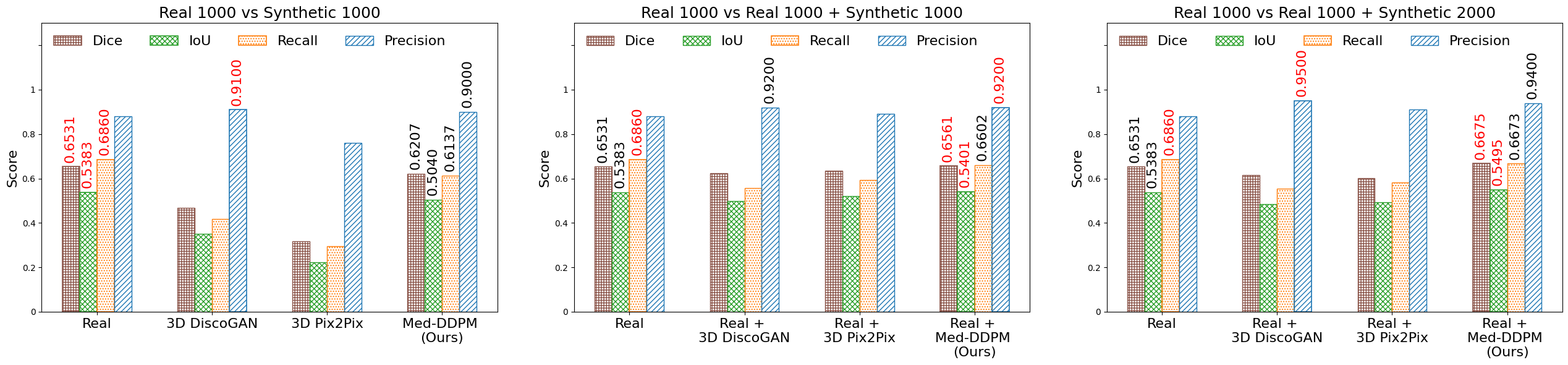}
\vspace{-6mm}
\caption{Comparative visualization of segmentation model outcomes across various methods, including baseline results for real images. Top performers are highlighted in red, and second-best results are marked in black.}
\label{fig6}
\end{figure*}

\subsection{Generated Images}
Fig.~\ref{fig2} presents coronal, sagittal, and axial slices of real brain MRI images alongside those generated by our proposed method, two conditional baseline models, and other unconditional 3D brain MRI synthesis models. This comparison underscores the differences in the overall quality of synthetic 3D brain MRIs produced by various methods. In terms of evaluating the overall quality of synthetic MRIs through visual assessment, the images generated by 3D StyleGAN appear blurry with wire mesh patterns. 3D-$\alpha$-WGAN-GP results in even blurrier images with similar textures, while HA-GAN produces the blurriest images, which are consistently asymmetric. LDM occasionally generates images with uniform textures and clearer edges. Although no model perfectly replicates continuous vessels, 3D-$\alpha$-WGAN-GP images exhibit better vessel continuity but with unnaturally wide vessels. The 3D Pix2Pix model generates blurry images, but 3D DiscoGAN offers better performance, creating more realistic images yet lacking detailed brain features and presenting coarse gyri and sulci. In contrast, our proposed Med-DDPM model produces images that are significantly more realistic than those from all other baseline methods in terms of the overall quality of 3D brain MRI synthesis.

Regarding our primary focus on conditional synthesis comparison, Fig.~\ref{fig3} provides a closer look at tumor areas in axial plane images, comparing real and synthetic images from Med-DDPM and two other conditional GAN models. The 3D Pix2Pix model struggles to the produce accurate images from unseen test masks, resulting in blurry outputs. In contrast, the 3D DiscoGAN model performs better than 3D Pix2Pix, generating more realistic images with clear tumor areas exhibiting complete ring enhancement and reduced blur compared to the 3D Pix2Pix model. However, it is worth noting that the 3D DiscoGAN generated images fail to clearly depict the distinctive features of the brain. Additionally, they generate coarse gyri and sulci, which appear unnatural. On the other hand, the proposed Med-DDPM model excels in generating highly realistic and high-quality images, with clear visibility of both brain features and tumor regions. Although it sometimes creates tumors with incomplete ring enhancement, this issue also occurs in real images. Furthermore, the peri-tumoral edema generated by Med-DDPM is more realistic, compared to the neatly isotropic low intensity generated by 3D DiscoGAN. These synthesized images bear a close resemblance to real ones, accurately capturing intricate details.
Fig.~\ref{fig4} showcases synthetic images created using manipulated masks, generated through two different functions: one that scales tumor masks from their center, and another that shifts them within the brain in axial, coronal, and sagittal planes. The 3D Pix2Pix model tends to create unrealistic tumor areas marked by noticeable artifacts. 3D DiscoGAN, on the other hand, produces clearer and more realistic tumor areas, yet it falls short in capturing detailed brain features, resulting in poor representations of gyri and sulci. Despite this, both models tend to produce similar brain features across images, suggesting a lack of diversity and indicating mode collapse. Conversely, Fig.~\ref{fig5} highlights the diversity in synthetic images generated from a single input mask, demonstrating that Med-DDPM is capable of producing a wide range of distinct samples.

\subsubsection{Quantitative Results}
The comprehensive quantitative evaluation of various generative models, including our proposed Med-DDPM, is detailed in Table~\ref{table:comparison}. Our Med-DDPM model showed exceptional performance across most metrics. It achieved the lowest MSE value of $0.0146$, highlighting its superior ability to preserve intricate details and structures. Additionally, with an MMD value of $28.2507$, Med-DDPM effectively matched the distribution of the target domain.

However, the 3D-FID score of Med-DDPM, at $144.8321$, was higher than that of most GAN models, suggesting that the feature extraction model might not be fully optimized for brain imaging features. Most notably, Med-DDPM achieved an MS-SSIM score of $0.6132$, the closest to the real data score of $0.5864$, underlining its excellence in maintaining structural integrity.

While models like 3D Pix2Pix and 3D StyleGAN excelled in specific metrics, with 3D Pix2Pix having the lowest MMD score of 24.9456 and 3D StyleGAN achieving the lowest 3D-FID score of 48.9729, Med-DDPM consistently maintained a balance across all metrics. This balanced performance underlines its overall effectiveness in medical image generation tasks, adeptly managing the diverse nature of the data.
\begin{table}[htbp]
\vspace{-4mm}
\centering
\caption{Quantitative Results}
\label{table:comparison}
\setlength{\tabcolsep}{3pt} 
\begin{tabular}{p{75pt}>{\raggedleft\arraybackslash}p{30pt}>{\raggedleft\arraybackslash}p{40pt}>{\raggedleft\arraybackslash}p{40pt}>{\raggedleft\arraybackslash}p{40pt}}
\hline 
Method & MSE$\downarrow$ & MMD$\downarrow$ & 3D-FID$\downarrow$ & MS-SSIM\\
\hline
3D-$\alpha$-WGAN-GP [21] & 0.0181 & 92.3998 & 74.5512 & 0.7610 \\
HA-GAN [22] & 0.0331 & 192.9155 & 1788.4518 & 0.4347\\
3D StyleGAN [23] & 0.0160 & 35.2398 & \textbf{48.9729} & 0.7429 \\
LDM [32] & 0.0349 & 330.3030 & 2730.7849 & 0.6584 \\
3D Pix2Pix [26] & 0.0171 & \textbf{24.9456} & 59.4183 & 0.6966 \\
3D DiscoGAN [41] & 0.0188 & 44.6890 & 86.3527 & 0.6730 \\
Med-DDPM (Ours) & \textbf{0.0146} & 28.2507 &  144.8321 & \textbf{0.6132} \\
\hline
Real & - & - & - & 0.5864\\
\hline
\end{tabular}
\vspace{-3.2mm}
\end{table}

\subsubsection{Qualitative results}
Regular visual assessment tests were conducted throughout the experimentation phase. The experts were presented with a mixture of real and synthetic 3D images generated by the proposed model and the baseline models. The experts evaluated the quality of the generated images. It was evident to the experts that the synthetic images produced by the baseline models exhibited blurriness, some artifacts and lacked realistic-looking brain features. In contrast, the synthetic images generated by our proposed method appeared more realistic. However, upon careful examination of the axial plane, the experts were able to identify the synthetic images due to slight inconsistencies in vessel continuity within the Circle of Willis area. Additionally, the synthetic images did not exhibit the presence of mass effects around large tumors, which typically result in shifts in the ventricles and the midline.

\begin{table*}[htbp]
\vspace{-5mm}
\centering
\caption{Performance Summary of Segmentation Models Trained on Synthetic Images} \vspace{2mm}
\label{table3}
\setlength{\tabcolsep}{12pt}
\begin{tabular}{llllll}
\hline
Experiment & Method                        & Dice          & IoU           & Recall        & Precision     \\
\hline
R 1000         & Real images                   & 0.6531±0.2831 & 0.5383±0.2565 & 0.6860±0.3159 & 0.8800±0.3250 \\
\hline
S 1000   
           & 3D DiscoGAN                      & 0.4685±0.2816 & 0.3497±0.2393 & 0.4169±0.2765 & \textbf{0.9100±0.2862} \\ 
           & 3D Pix2Pix & 0.3171±0.2706 & 0.2219±0.2094 & 0.2957±0.2759 & 0.7600±0.4271 \\
    & \textbf{Med-DDPM (Ours)} & \textbf{0.6207±0.2882} & \textbf{0.5040±0.2610} & \textbf{0.6137±0.3156} & 0.9000±0.3000 \\
\hline
R 500 + S 500   
& 3D DiscoGAN & 0.6098±0.2775 & 0.4888±0.2534 & 0.5458±0.2818 & 0.9200±0.2713 \\
& 3D Pix2Pix & 0.6135±0.2989 & 0.5011±0.2742 & 0.5836±0.3161 & 0.9100±0.2862 \\
& \textbf{Med-DDPM (Ours)} & \textbf{0.6449±0.2769} & \textbf{0.5272±0.2534} & \textbf{0.6832±0.3125} & \textbf{0.9300±0.2551} \\
\hline
R 1000 + S 1000  
& 3D DiscoGAN & 0.6239±0.2627 & 0.4989±0.2404 & 0.5570±0.2678 & 0.9200±0.2713 \\
          & 3D Pix2Pix & 0.6343±0.2928 & 0.5211±0.2670 & 0.5938±0.3010 & 0.8900±0.3129 \\
& \textbf{Med-DDPM (Ours)} & \textbf{0.6561±0.2758} & \textbf{0.5401±0.2553} & \textbf{0.6602±0.3028} & \textbf{0.9200±0.2713} \\
\hline
R 1000 + S 2000  
           & 3D DiscoGAN & 0.6141±0.2543 & 0.4848±0.2295 & 0.5548±0.2537 & \textbf{0.9500±0.2179}\\
           & 3D Pix2Pix & 0.6010±0.3544 & 0.4915±0.2030 & 0.5820±0.2895 & 0.9100±0.2713 \\
           & \textbf{Med-DDPM (Ours)} & \textbf{0.6675±0.2623} & \textbf{0.5495±0.2489} & \textbf{0.6673±0.2949} & 0.9400±0.2375 \\
\hline
\multicolumn{6}{p{450pt}}{Evaluation results for a segmentation model trained with synthetic images. 'R' stands for real images, the baseline; 'S' for synthetic images used in training; and 'R + S' for a mix of both. Scores are in mean ± standard deviation format, with the best results highlighted in bold compared to the baseline.}\\
\end{tabular}
\end{table*}

\begin{figure*}[htbp]
\centering
\includegraphics[width=1.0\textwidth]{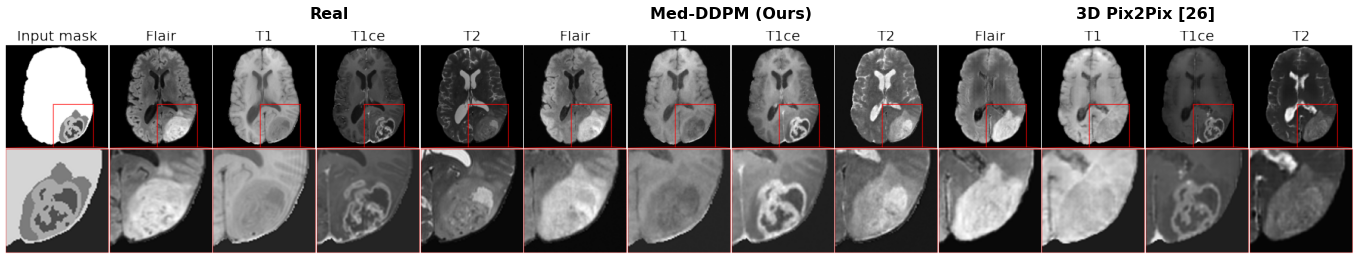}
\vspace{-6mm}
\caption{Axial view comparison of real and generated images in the 4 modalities synthesis experiment (Largest tumor slice)}
\label{fig7}
\end{figure*}
\begin{table*}[htbp]
\centering
\caption{Performance Comparison of Memory Consumption and Training Speed}
\label{table_memory_speed}
\begin{tabular}{llllllll}
\hline
\# & Model & \multicolumn{3}{l}{Med-DDPM (Ours)} & \multicolumn{3}{l}{3D Pix2Pix [21]} \\
& & Training & Inference & Training Speed & Training & Inference & Training Speed \\
& & (mb) & (mb) & (iter/s) & (mb) & (mb) & (iter/s) \\
\hline
1 & Single-modality synthesis  & 23981 & 7991 & 0.65 & 17138 & 7046 & 13.21 \\
2 & Multi-modality synthesis  & 80963 & 21159 & 0.61 & 69225 & 18299 & 5.89 \\
\hline
\end{tabular}
\vspace{-4mm}
\end{table*}
\subsection{Comparison of Segmentation Models Trained on Synthetic Images}
The evaluation presented in Table~\ref{table3} highlights the performance of segmentation models trained on synthetic images. Fig.~\ref{fig6} visually compares these results, clearly indicating that our Med-DDPM model outperforms the baseline models, 3D DiscoGAN and 3D Pix2Pix.

In scenarios involving 1,000 real images, 1,000 synthetic images, and their combinations, Med-DDPM consistently outperformed the baseline models. Specifically, in the experiment with solely 1,000 synthetic images, Med-DDPM achieved a Dice score of 0.6207, surpassing 3D DiscoGAN (0.4685) and 3D Pix2Pix (0.3171). In combined scenarios, such as 1,000 real images with 1,000 synthetic images, Med-DDPM maintained its lead with a Dice score of 0.6561, compared to 0.6239 for 3D DiscoGAN and 0.6343 for 3D Pix2Pix.

Moreover, in the larger dataset comprising 1,000 real images and 2,000 synthetic images, the performance of Med-DDPM reached its peak with a Dice score of 0.6675, surpassing the baseline score of 0.6531 for real images and demonstrating its potential for data augmentation capabilities.

While 3D DiscoGAN and 3D Pix2Pix models demonstrated improvements in mixed data scenarios, they were consistently outclassed by Med-DDPM. Only 3D DiscoGAN achieved the highest precision score of 0.91 on 1,000 synthetic images, where the real precision score was 0.88. Additionally, it attained a precision score of 0.95 on combinations of 1,000 real and 2,000 synthetic images. This implies that the model has a low rate of false positives in its predictions and is highly proficient at accurately delineating the boundaries or regions corresponding to the positive class (i.e., the tumor class) in the segmentation task. Our proposed Med-DDPM also achieved a high precision score, closely approaching that of the 3D DiscoGAN.

\subsection{3D Multimodal MRI Synthesis Experiment}
In this section, we present the results of an additional experiment conducted to validate the effectiveness of our proposed method, Med-DDPM, for multimodal MRI synthesis. We utilized the brain-extracted MRI dataset from the BraTS2021 challenge\footnote{http://braintumorsegmentation.org/} for this experiment, with the objective of demonstrating the capability of our method in generating high-quality images for all four MRI modalities (T1, T1CE, T2, and Flair) simultaneously from a segmentation mask.

To enhance model training efficiency, we selected 193 high-quality images where all modalities have no distortion and artifacts from the dataset and preprocessed them by applying cropping and padding to a size of 192x192x144. The original labels of the segmentation mask in this dataset are 0, 1, 2, and 4, where 1, 2, and 4 represent tumor parts. For better adaptation to our needs, we modified the mask labels as follows: we changed mask labels from 4 to 3 and introduced one more label (label 4) to represent the brain area, achieved by thresholding the T1 image. Consequently, the final class labels were defined as follows: 0 represents the background, 1 to 3 correspond to the tumor parts, and 4 indicates the brain area.

Next, we performed a one-hot encoding operation, excluding the background class label 0. This operation resulted in a mask image with four channels, each representing one of the classes. During the training process, we applied channel-wise concatenation to combine the input training image (including all four modalities T1, T1CE, T2, and Flair concatenated together, resulting in a four-channel image) with the final mask image, also a four-channel image. The resulting concatenated image contains eight channels, and this configuration was used in the training process. For the Med-DDPM model in this experiment, we used the same experimental setups as in our main model training, except we employed 200,000 iterations for this experiment. Due to the high resolution nature of the eight-channel image, we trained our model on the NVIDIA A100 80GB GPU card for 5 days. 

In this experiment, we also employed C.-H.Shin et al.\cite{nvidia} as our comparison baseline model since originally this paper is only one study in the literature investigating 3D semantic brain MRI synthesis especifically 4 modalities synthesis.  

Fig.~\ref{fig7} illustrates a comparison between the generated samples from our experiments and the corresponding real images. The images generated by our proposed method exhibit a high level of fidelity and closely resemble the actual MRI modalities than baseline GAN model. This outcome further highlights the robustness and accuracy of our approach in effectively generating multiple modalities from a single segmentation mask.
\subsection{Memory Efficiency}
In this section, we compared the memory efficiency of our model with the baseline model by measuring GPU memory usage and training speed. We adopted the training speed measurement methodology used by L. Sun et al. \cite{hagan}, which involved recording the number of iterations per second during training. For the training of multi-modality synthesis, we employed an NVIDIA Tesla A100 GPU, and for single modality synthesis, a Tesla V100–SXM2 32 GB GPU card was used. Both models were trained with a batch size of 1. Table~\ref{table_memory_speed} provides the detailed comparison of memory consumption and training speed. Although our proposed method produced more realistic-looking images than the baseline model, the architecture of our diffusion model necessitated higher memory consumption compared to the baseline 3D pix2pix GAN model.

\section{Discussion}
The findings of this study underscored the significant advancements in 3D brain MRI synthesis using generative models. The Med-DDPM model, in particular, demonstrated remarkable capabilities in generating realistic and detailed brain images, marking a critical advancement in medical imaging.

In the comparison of various generative models for evaluating the overall quality of synthetic brain MRIs (Fig.~\ref{fig2}), Med-DDPM consistently outperformed baseline models such as 3D StyleGAN, HA-GAN, LDM, and 3D-$\alpha$-WGAN-GP, especially in maintaining the structural integrity and realistic representation of both normal brain tissue and pathological features like tumors. This was evident from the quantitative results presented in Table~\ref{table:comparison}, where Med-DDPM achieved the lowest MSE score of 0.0146 and was closest to the real image, with an MS-SSIM score of 0.6132 compared to the MS-SSIM score of 0.5864 for real images. However, its higher 3D-FID score, compared to other models, suggests that there is room for further optimization in feature extraction specific to brain imaging.

Furthermore, the qualitative assessments by experts validated the superiority of the Med-DDPM model over other baseline conditional models, such as 3D Pix2Pix and 3D DiscoGAN, in generating synthetic images that closely resemble real brain MRIs. These evaluations highlighted the potential of Med-DDPM to replicate the intricate details of brain anatomy and pathology with remarkable fidelity. However, the minor inconsistencies noted in vessel continuity and the lack of mass effects around large tumors highlight areas for future improvement.

The evaluations of segmentation models (Table~\ref{table3}) further reinforce the utility of synthetic images generated by Med-DDPM, particularly in data augmentation. The superior performance of the model in datasets with a combination of real and synthetic images demonstrates its potential in enhancing the training of segmentation algorithms, thus contributing significantly to advancements in medical image analysis.
\section{Conclusion}
This study introduces Med-DDPM, a benchmark for generating 3D semantic brain MR images, filling a significant gap in the literature. Our method is a conditional diffusion model that directly transforms random noise of the same dimension as the output image into realistic images by conditioning on the segmentation mask. This approach differs significantly from other existing methods, which employ latent vectors as inputs to generate images. The unique semantic conditioning of our model enables the generation of diverse and anatomically accurate images, setting a new standard in image fidelity and opening up new possibilities for image anonymization. The effectiveness of Med-DDPM is demonstrated by its performance in tumor segmentation tasks and its ability to generate all four MRI modalities from a single segmentation mask. This versatility is crucial for accurately representing complex brain structures. Overall, Med-DDPM not only showcases the capabilities of diffusion models in generating high-quality medical images but also addresses challenges such as data scarcity and privacy in the healthcare domain. Future work should focus on expanding its applications and refining its capabilities to further advance medical imaging.

\section*{Acknowledgment}
This work was supported by the National Science and Technology Council, Taiwan [Grant No. 111-2221-E-002-049-MY3, 112-2221-E-011-111, NSTC 112-2634-F-011-002-MBK] and National Taiwan University Hospital [Grant No. 110-EDN03].

\end{document}